\documentclass[9pt,conference]{IEEEtran}
\IEEEoverridecommandlockouts
\usepackage{cite}
\usepackage{amsmath,amssymb,amsfonts}
\usepackage{algorithmic}
\usepackage{graphicx}
\usepackage{textcomp}
\usepackage{xcolor}
\usepackage{hyperref}
\usepackage{colortbl}
\usepackage{multirow}
\usepackage{footnote}
\usepackage{tablefootnote}
\usepackage{float}
\usepackage{threeparttable}
\usepackage{booktabs}
\usepackage{multicol}
\usepackage{afterpage}
\usepackage{stfloats}

\newcommand{\centercaption}[1]{\caption{\centering #1}}

\newcommand{\method}{UniCAIM}

\hypersetup{
    colorlinks=true,
    linkcolor=blue,
    filecolor=magenta,      
    urlcolor=cyan,
    pdftitle={Overleaf Example},
    citecolor=blue,
}

\def\BibTeX{{\rm B\kern-.05em{\sc i\kern-.025em b}\kern-.08em
    T\kern-.1667em\lower.7ex\hbox{E}\kern-.125emX}}
    
\definecolor{Gray}{gray}{0.85}

\begin{document}
\title {UniCAIM: A Unified CAM/CIM Architecture with Static-Dynamic KV Cache Pruning for Efficient Long-Context LLM Inference \vspace{-10pt}}


\author{\\}
  \author{
      Weikai Xu$^{1\#}$,
      Wenxuan Zeng$^{42\#}$, Qianqian Huang$^{13}$, Meng Li$^{213*}$, and Ru Huang$^{13*}$
  \\
  \textit{$^1$School of Integrated Circuits, Peking University, China;} 
  \textit{$^2$Institute for Artificial Intelligence, Peking University, China;} \\
  \textit{$^3$Beijing Advanced Innovation Center for Integrated Circuits, Beijing, China;} \\
  \textit{$^4$School of Software and Microelectronics, Peking University, China.} 
  \textit{$^{\#}$Equal contribution.} \\ \vspace{-20pt}

\thanks{
This work was supported in part by NSFC under Grant 62495102 and Grant 92464104, in part by the National Key Research and Development Program under Grant 2024YFB4505004, in part by Beijing Municipal Science and Technology Program under Grant Z241100004224015, and in part by 111 Project under Grant B18001.

$^*$Corresponding author: ruhuang@pku.edu.cn; meng.li@pku.edu.cn}}

\maketitle

\begin{abstract}
Transformer-based large language models (LLMs) have achieved impressive performance in various natural language processing (NLP) applications. However, the high memory and computation cost induced by the KV cache limits the inference efficiency, especially for long input sequences. Compute-in-memory (CIM)-based accelerators have been proposed for LLM acceleration with KV cache pruning. However, as existing accelerators only support static pruning with a fixed pattern or dynamic pruning with primitive implementations, they suffer from either high accuracy degradation or low efficiency. In this paper, we propose a ferroelectric FET (FeFET)-based unified content addressable memory (CAM) and CIM architecture, dubbed as \method. \method~features simultaneous support for static and dynamic pruning with 3 computation modes: 1) in the CAM mode, \method~enables approximate similarity measurement in $\mathcal{O}(1)$ time for dynamic KV cache pruning with high energy efficiency; 2) in the charge-domain CIM mode, static pruning can be supported based on accumulative similarity score, which is much more flexible compared to fixed patterns; 3) in the current-domain mode, exact attention computation can be conducted with a subset of selected KV cache. We further propose a novel CAM/CIM cell design that leverages the multi-level characteristics of FeFETs for signed multi-bit storage of the KV cache and in-place attention computation. With extensive experimental results, we demonstrate \method~can reduce the area-energy-delay product (AEDP) by 8.2$\sim$831$\times$ over the state-of-the-art CIM-based LLM accelerators at the circuit level, along with high accuracy comparable with dense attention at the application level, showing its great potential for efficient long-context LLM inference.

\end{abstract}
\section{Introduction}
\label{sec:intro}

\begin{figure}[tb]
    \centering
    \includegraphics[width=1\linewidth]
    {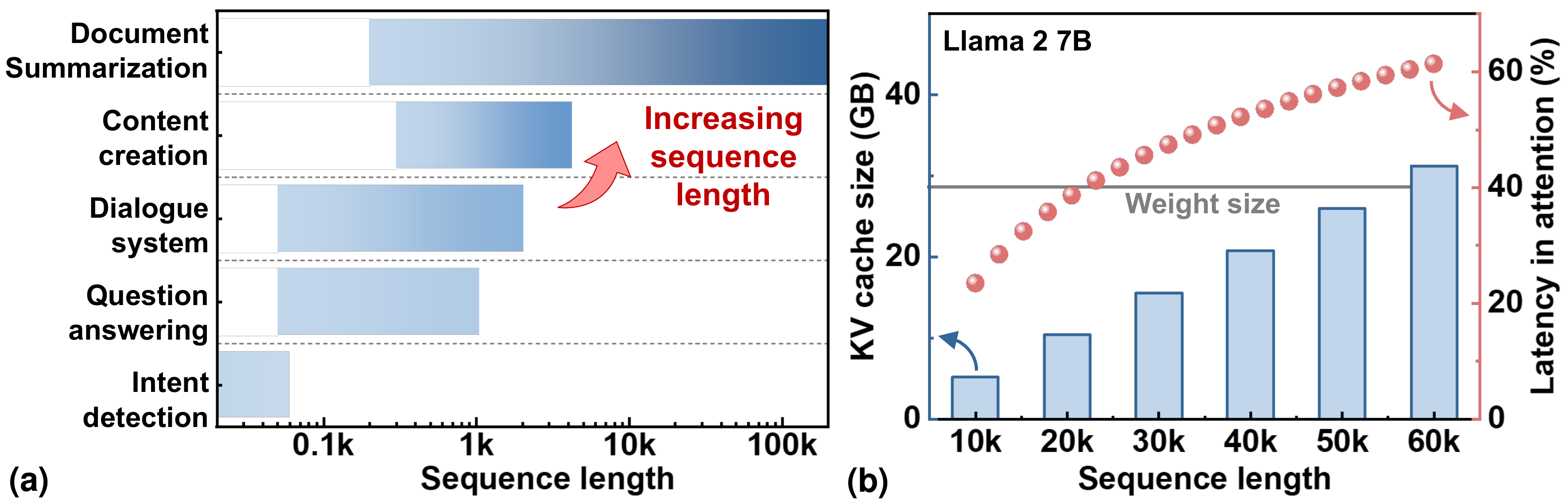}
    \vspace{-20pt}
    \caption{(a) Various NLP tasks with increasing sequence length. (b) The impact of sequence length on KV cache size and attention latency in Llama-2-7B, which is a typical LLM, indicating the memory and computation challenges faced by long-context LLMs.}
    \vspace{-18pt}
    \label{fig:1}
\end{figure}

Large language models (LLMs) have recently demonstrated remarkable performance in a wide range of
applications such as question answering, code completion, and dialogue systems \cite{yi2024survey,sudhakaran2024mariogpt,van2024adapted}. The context length supported by LLMs is also growing progressively to support more applications like multi-turn chat, text summarization, etc. However, with the increase in sequence lengths, the KV cache size gradually exceeds the LLM parameter size and the attention computation latency is becoming increasingly dominant, both of which emerge as prominent bottlenecks in long-context LLM inference \cite{luohe2024keep}, as shown in Fig. \ref{fig:1}.


Recently, various solutions have been proposed to reduce the KV cache overhead from the algorithm perspective, utilizing the highly sparse nature of attention \cite{zhao2024alisa,tang2024quest}. There are static and dynamic KV cache pruning policies, aiming to either reduce memory footprint by permanently evicting unnecessary tokens or reduce the computation load by only fetching the KV pairs with high attention scores, respectively \cite{zhang2024h2o,zhao2024alisa,tang2024quest,li2024snapkv}. Meanwhile, from the hardware perspective, the computing-in-memory (CIM) architecture which can perform the general matrix-vector multiplication (GEMV) operations within the memory array, has been proven to compute attention efficiently by reducing the data movement \cite{chen2018efficient, xu2024heirs, lu2023rram, yang2020retransformer}. Besides, some works specifically adopt KV cache pruning policies with CIM architecture to further support sparse attention computation \cite{trancim, tu202316, cimformer, tu2023161, yazdanbakhsh2022sparse, zheng2023accelerating}.

However, the current CIM-based LLM accelerators only focus on either static or dynamic KV cache pruning algorithms, hindering the simultaneous optimization of memory usage and computation overhead.
Moreover, the existing hardware implementations are relatively primitive, suffering from the following challenges. On the one hand, the current CIM designs 
adopting static KV cache pruning only support fixed sparse attention pattern \cite{trancim, tu202316}, leading to accuracy degradation, especially for long sequences \cite{xiao2023efficient,beltagy2020longformer}. Although there are advanced token-aware static KV pruning policies with better accuracy, they require the computation of accumulative attention scores, which will lead to a significant amount of extra computation and power consumption in current CIM architectures. On the other hand, to further support dynamic KV cache pruning, additional expensive operators such as top-\textit{k} selection, are required, suffering from high hardware cost, latency, and energy consumption \cite{cimformer, tu2023161}.
Therefore, it is challenging for the current CIM architectures to efficiently support both dynamic and static KV cache pruning simultaneously, restricting the practical deployment of LLMs with increasing sequence length.

In this work, a unified content addressable memory (CAM) and CIM architecture called UniCAIM, featuring static-dynamic KV cache pruning for long-context LLM inference, is proposed through algorithm-hardware co-optimization.
UniCAIM addresses the challenges mentioned above, and the main contributions can be summarized as follows:
\begin{itemize}
    \item A hardware-friendly static-dynamic KV cache pruning framework is proposed, which can significantly improve the inference efficiency of LLMs without degrading the model. 
    Firstly, the unimportant tokens are dropped permanently in the prefill stage, reducing the overall memory overhead.
    Moreover, during each decoding step, only the most relevant (i.e., top-\textit{k}) tokens are selected for exact attention computation to reduce the computation load, and one token is evicted based on accumulative attention scores to enhance memory utilization and management when the generated length exceeds the reserved KV cache size.
    \item A UniCAIM architecture with CAM and CIM modes is proposed for efficiently implementing the static-dynamic KV cache pruning and sparse attention computing.
    Based on the CAM mode of UniCAIM, attention scores can be approximately measured and compared to select the top-\textit{k} tokens in $\mathcal{O}(1)$ time without the exact computations, and the attention scores can also be accumulated for static eviction by further leveraging the charge-domain CIM within the same operation cycle, leading to fast and energy-efficient static-dynamic KV cache pruning.
    Moreover, the current-based CIM is utilized for the exact attention score computation of selected tokens with high precision.
    \item The proposed UniCAIM is implemented based on emerging compact ferroelectric FET (FeFET) devices with carefully designed circuits in hardware for further area-energy-delay product (AEDP) reduction.
    Experiments and evaluations show that the proposed FeFET-based UniCAIM can achieve 8.2$\sim$831$\times$ AEDP reduction compared with the state-of-the-art CIM-based LLM accelerators, showing its great potential for high speed, area- and energy-efficient long-context LLM acceleration.
\end{itemize}

\section{Background}
\label{sec:background}

In this section, we review the sparse attention in transformer-based LLMs and exiting KV cache pruning policies from the algorithm and hardware perspectives, as well as the basics of FeFET and its advantages for CIM and CAM.


\subsection{Sparse attention and KV cache pruning} 
\subsubsection{\textbf{Sparse attention and existing KV cache pruning algorithms}}

In transformers, the quadratic computational complexity of attention is one of the major bottlenecks \cite{vaswani2017attention}.
Many recent research efforts have been devoted to exploiting the intrinsic sparsity in attention \cite{jiang2024minference}.
For long-context generation tasks, KV cache pruning becomes a promising solution for efficient sparse attention computation.
Existing KV cache pruning methods can be roughly categorized into static pruning and dynamic pruning.
Some static pruning policies predefine the positions of important tokens and remain consistent across decoding steps, such as StreamingLLM \cite{xiao2023efficient}, while such fixed patterns lack flexibility for different LLMs and contexts.
Other static pruning policies permanently drop some tokens and the dropped ones cannot be used in the subsequent decoding steps, such as SnapKV \cite{li2024snapkv} and H2O \cite{zhang2024h2o}. However, only tokens that remain throughout the entire decoding process can be pruned, otherwise suffering significant accuracy degradation.
On the other hand, dynamic pruning policies select unimportant tokens to drop at for different decoding steps, such as InfLLM \cite{xiao2024infllm} and LongCache \cite{liu2024farewell}. Though dynamic pruning is more flexible than static pruning, it involves more expensive computations, such as attention score ranking and top-\textit{k} selection, which pose greater challenges for efficient hardware implementation.

\subsubsection{\textbf{Existing hardware implementations for KV cache pruning}} 
KV cache pruning in LLMs offers significant advantages by reducing memory usage and redundant computation. 
Recently, CIM-based transformer accelerators have attracted widespread attention for LLMs due to the largely reduced data movement overhead, with some studies exploring the implementation of KV cache pruning in hardware. 
On the one hand, some works implement fix-pattern KV cache pruning which is well-suited for CIM architecture, such as TranCIM \cite{trancim} adopting the pruning algorithm proposed in StreamingLLM \cite{xiao2023efficient}. 
However, this approach performs computations on neighboring tokens within a predetermined fixed attention range defined in the algorithm, which lacks flexibility and fails to consider the varying contributions of different tokes to the final prediction. 
Consequently, the CIM with fix-pattern KV cache pruning designs suffers from suboptimal accuracy and efficiency, especially when processing different LLMs and contexts. 
On the other hand, some works have implemented dynamic KV cache pruning, primarily involving attention score ranking and selection, such as CIMFormer \cite{cimformer} adopting the top-\textit{k} selection. 
However, it requires additional hardware overhead to achieve top-\textit{k} selection with complex time complexity of $\mathcal{O}(nlogn)$ \cite{wang2021spatten} or $\mathcal{O}(logn)$ with additional circuits for token gathering \cite{cimformer}, resulting in high latency and power consumption. 
Besides, there are emerging non-volatile memories (NVMs)-based CIM designs for dynamic pruning by utilizing approximate attention scores \cite{yazdanbakhsh2022sparse, zheng2023accelerating}, but suffering the trade-off between energy efficiency and accuracy.

Therefore, current CIM-based LLM accelerators support either static or dynamic KV cache pruning policies, along with relatively primitive implementations, facing substantial penalties in terms of area, latency, and power consumption (Table \ref{table:1}). 
Thus, while CIM-based KV cache optimization techniques bring new possibilities for efficient inference of long-context LLMs, there are still several challenges that need to be addressed. 

\subsection{FeFET for CIM and CAM} 
\begin{figure}[tb]
    \centering
    \includegraphics[width=1\linewidth]
    {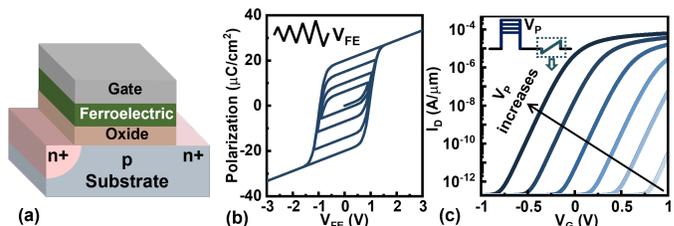}
    \vspace{-24pt}
    \caption{(a) Typical device structure of ferroelectric FET (FeFET). (b) FE polarization-voltage loops with multilevel FE polarizations. (c) Gradually modulated I\textsubscript{D}-V\textsubscript{G} curves of FeFET for the multilevel storage capability.}
    \vspace{-15pt}
    \label{fig:2}
\end{figure}

{\renewcommand{\thefigure}{\Roman{figure}}
\begin{figure}[tb]
    \renewcommand{\figurename}{Table} 
    \setcounter{figure}{0} 
    \centercaption{Qualitative comparison of proposed FeFET-based UniCAIM with the state-of-the-art CIM-based LLM accelerators.}
    \vspace{3pt}
    \centering
    \includegraphics[width=1\linewidth]
    {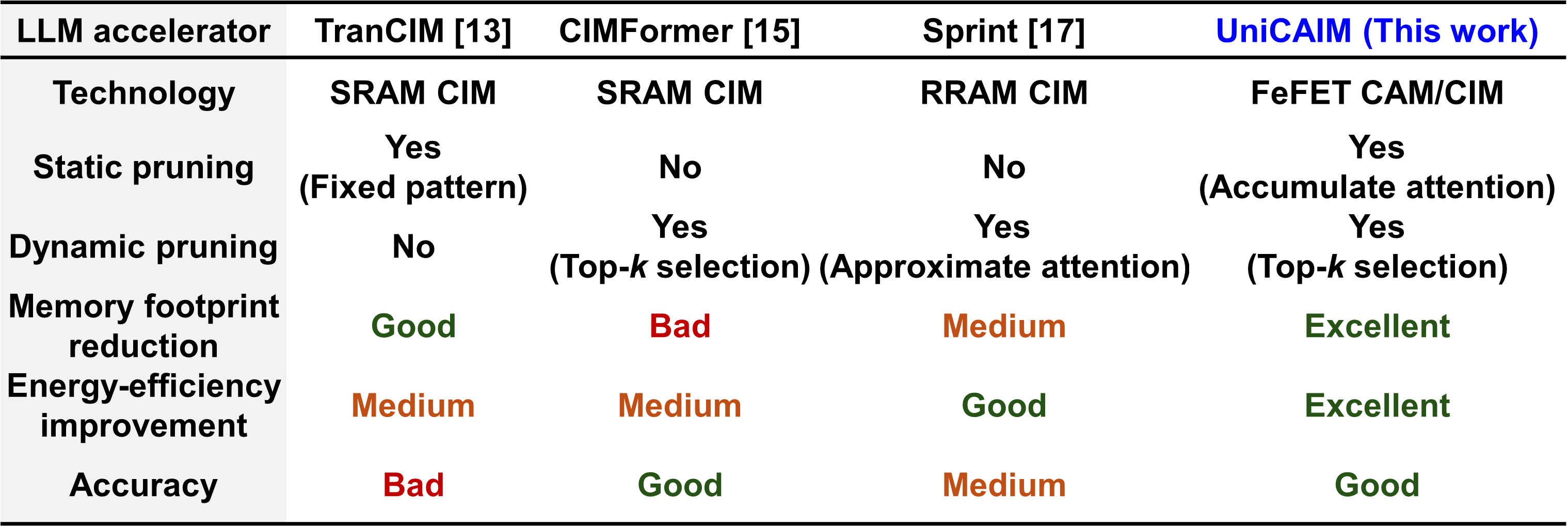}
    \vspace{-28pt}
    \label{table:1}
\end{figure}
}
In recent years, various emerging NVMs, such as resistive random-access memory (RRAM), magnetic tunnel junction (MTJ) and FeFET, have triggered lots of attention for CIM, due to the high storage density and efficient GEMV operation via analog computing within the memory array \cite{sebastian2020memory,yin2021enabling,frank2023impact}. 
On the other hand, CAM is a specialized memory architecture, where an input query is compared with entire stored entries simultaneously within the memory array \cite{pagiamtzis2006content,xu2024compact,hu2021memory}. Different from conventional CIM, CAM only identifies the matching entry or measures the matching degree without precise quantification, which is more energy efficient.
Among the NVMs, the three-terminal FeFET has the advantages of low write energy due to the electric-field-driven write mechanism and high I\textsubscript{ON}/I\textsubscript{OFF} ratio due to the transistor structure, along with good CMOS compatibility \cite{dutta2022logic}. 
Fig. \ref{fig:2}(a) shows the structure of an FeFET device, with a HfO\textsubscript{2}-based ferroelectric (FE) layer integrated into the gate stack of a MOSFET. 
Applying different program voltages (V\textsubscript{P}) to the gate of FeFET will switch the FE polarization states (Fig. \ref{fig:2}(b)), and thus gradually modulate the threshold voltage (V\textsubscript{TH}) of FeFET, indicating the multilevel storage capability (Fig. \ref{fig:2}(c)). 
Besides, by applying a relatively small read voltage (V\textsubscript{R}), the channel conductance state can be non-destructively readout without FE polarization switching. Therefore, FeFET can achieve both storage and computation, making it a promising candidate for CIM and CAM designs \cite{yin2021enabling, hu2021memory}.


\section{UniCAIM Architecture with Static-Dynamic KV Cache Pruning}
\label{sec:core}

\subsection{Hybrid Static-Dynamic KV Cache Pruning Algorithm}


\begin{figure*}[tb]
    \centering
    \setcounter{figure}{2} 
    \includegraphics[width=1\linewidth]
    {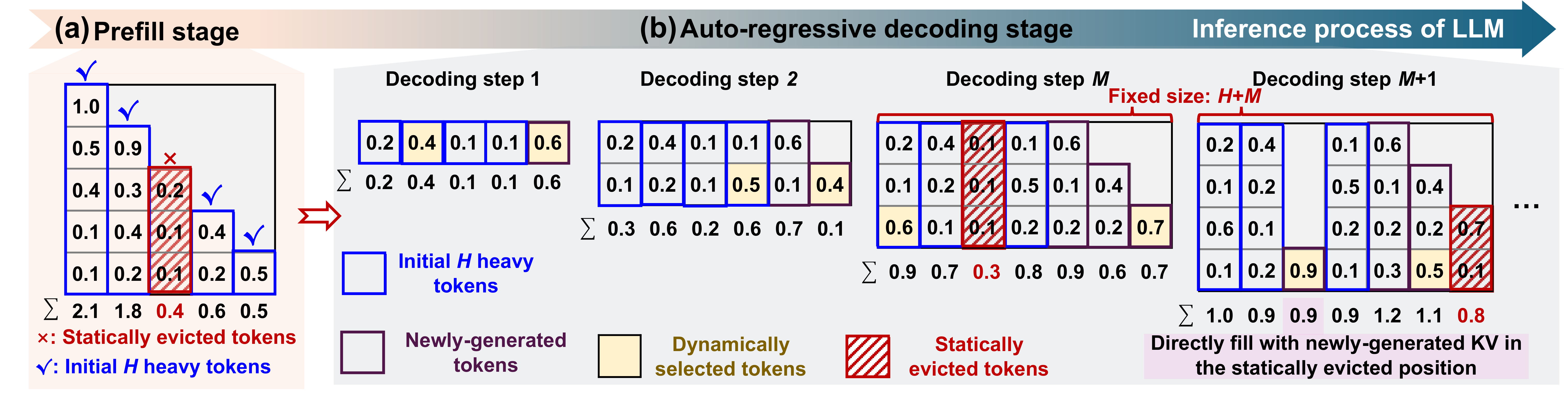}
    \vspace{-20pt}
    \caption{Framework of the proposed hybrid static-dynamic KV cache pruning algorithm. (a) During the prefill stage, static pruning evicts unimportant tokens for the subsequent generation. (b) During the decoding stage, dynamic pruning preserves a subset of tokens for sparse attention computation, while static pruning evicts one token at each step when the generated length exceeds the reserved size for a fixed KV cache size.}
    \vspace{-10pt}
    \label{fig:3}
\end{figure*}

\begin{figure*}[tb]
    \centering
    \includegraphics[width=1\linewidth]
    {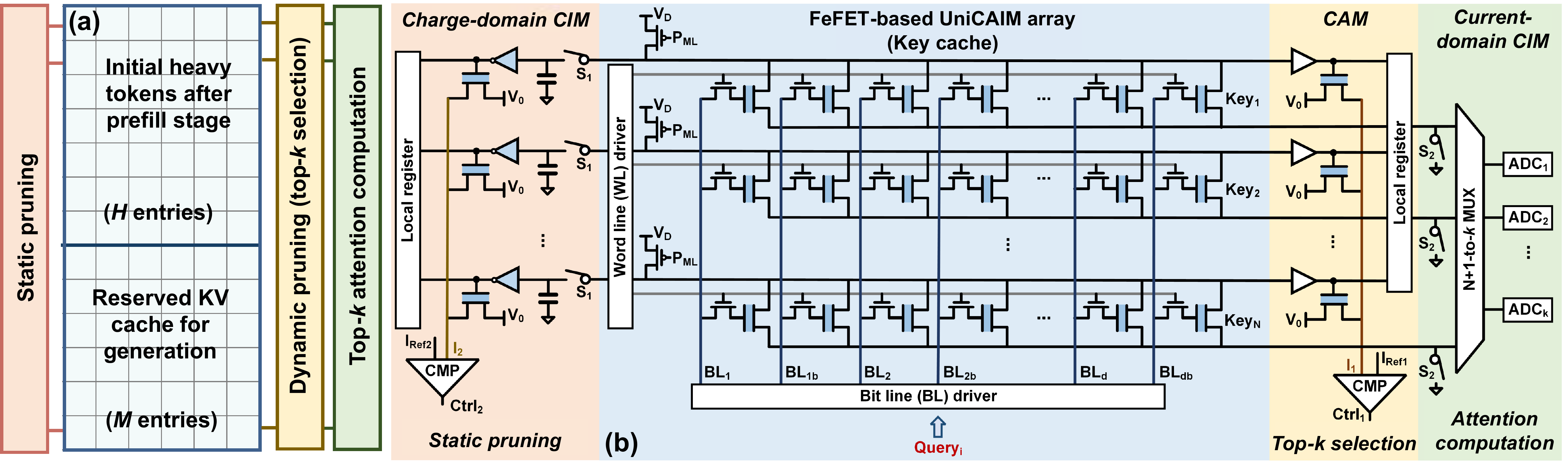}
    \vspace{-20pt}
    \caption{(a) The proposed UniCAIM architecture for static-dynamic KV cache pruning and sparse attention computing. (b) The hardware design of UniCAIM based on FeFET, including FeFET-based UniCAIM array and carefully designed peripheral circuits for CAM, charge-domain and current-domain CAM.}
    \vspace{-15pt}
    \label{fig:4}
\end{figure*}

The framework of the proposed hybrid pruning algorithm is illustrated in Fig. \ref{fig:3}, which combines static pruning during the prefill stage for an overall memory footprint reduction and static-dynamic pruning during the decoding stage for further efficient sparse attention.

\subsubsection{\textbf{Prefill stage with one-shot static pruning}}
During the prefill stage, we leverage static pruning to drop a part of tokens permanently which are almost unimportant throughout the entire decoding process, and remain the heavy tokens.
Following \cite{zhang2024h2o,zhao2024alisa,li2024snapkv}, we determine which tokens to statically drop based on the accumulative attention scores as shown in Fig. \ref{fig:3}(a).
Specifically, we drop the tokens with the lowest accumulated attention scores, reducing the overhead of the entire inference process.

\subsubsection{\textbf{Decoding stage with step-wise static-dynamic pruning}}
Since the LLM attention exhibits high sparsity \cite{zhao2024alisa} and different queries can attend to different tokens \cite{tang2024quest},
we propose to further selectively choose a subset of important tokens for efficient attention computation at different decoding steps (Fig. \ref{fig:3}(b)).
We adopt the most commonly used Cosine similarity as the attention score ($Attn$) to evaluate the importance of the KV cache for previous tokens as follows:
\begin{align}
    Attn(q, K) = q \cdot K^{T},
\end{align}
where $ q\in \mathbb R^{h\times 1\times d}$ denotes the query at the current step, $K\in \mathbb R^{h\times N\times d}$ denotes the key cache, and $h$, $N$, $d$ represent the number of heads, the number of tokens in key cache, and hidden dimension, respectively.
After the similarity measure, we leverage the top-\textit{k} selection strategy to pick up the most important tokens for the subsequent sparse attention. 

Moreover, when considering the memory limitation and hardware constraints, the consistently increasing KV cache size during the decoding process is not friendly.
To address this problem, we always keep a table of the accumulated attention scores and statically drop one token with the lowest accumulated attention score when the generated length exceeds the reserved size for decoding, and directly fill with the newly-generated token in the statically evicted position. As a result, the KV cache size is fixed with enhanced memory utilization and management.

\subsection{FeFET-based UniCAIM Architecture}
\subsubsection{\textbf{Overview of FeFET-based UniCAIM}}
Fig. \ref{fig:4}(a) shows the proposed UniCAIM architecture for static-dynamic KV cache pruning, which is specially optimized for the auto-regressive decoding stage, due to the memory-bound nature of the decoding stage. After one-shot static pruning at the prefill stage, \textit{H} heavy tokens with the highest accumulated attention scores are retained and stored in the UniCAIM array, and \textit{M} entries are reserved for newly generated tokens at the decoding stage. During each encoding step, the approximate attention scores are evaluated for dynamically selecting the top-\textit{k} tokens with the highest scores, which will be accurately computed. Additionally, the accumulated attention scores are also evaluated for statically evicted the token with the lowest scores, when the number of generated tokens exceeds \textit{M}. The step-wise static pruning during the decoding stage can maintain a fixed-size KV cache, resulting in higher area efficiency and better memory utilization. In this work, the proposed UniCAIM architecture is implemented using emerging FeFET in hardware (Fig. \ref{fig:4}(b)), incorporating an FeFET-based UniCAIM array which is shared by CAM and CIM, along with carefully designed peripheral circuits dedicated to dynamic pruning, static pruning, and attention computation.

\begin{figure}[tb]
    \centering
    \includegraphics[width=1\linewidth]
    {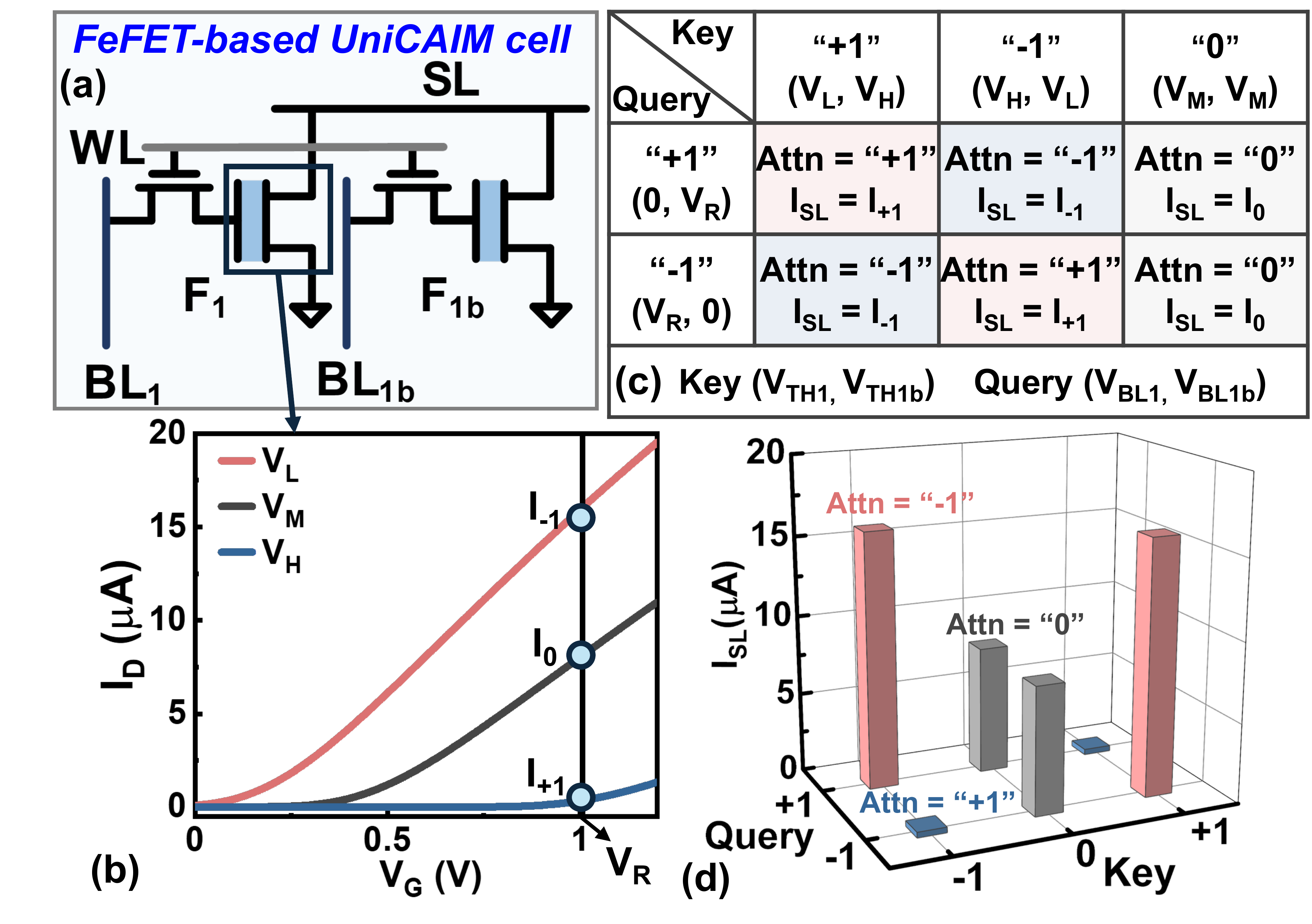}
    \vspace{-22pt}
    \caption{(a) The proposed FeFET-based UniCAIM cell. (b) Modulated threshold voltages (V\textsubscript{TH}) of FeFET. (c) The truth table of signed key and query. (d) The computing results of local signed multiplication.}
    \vspace{-10pt}
    \label{fig:5}
\end{figure}

\begin{figure}[tb]
    \centering
    \includegraphics[width=1\linewidth]
    {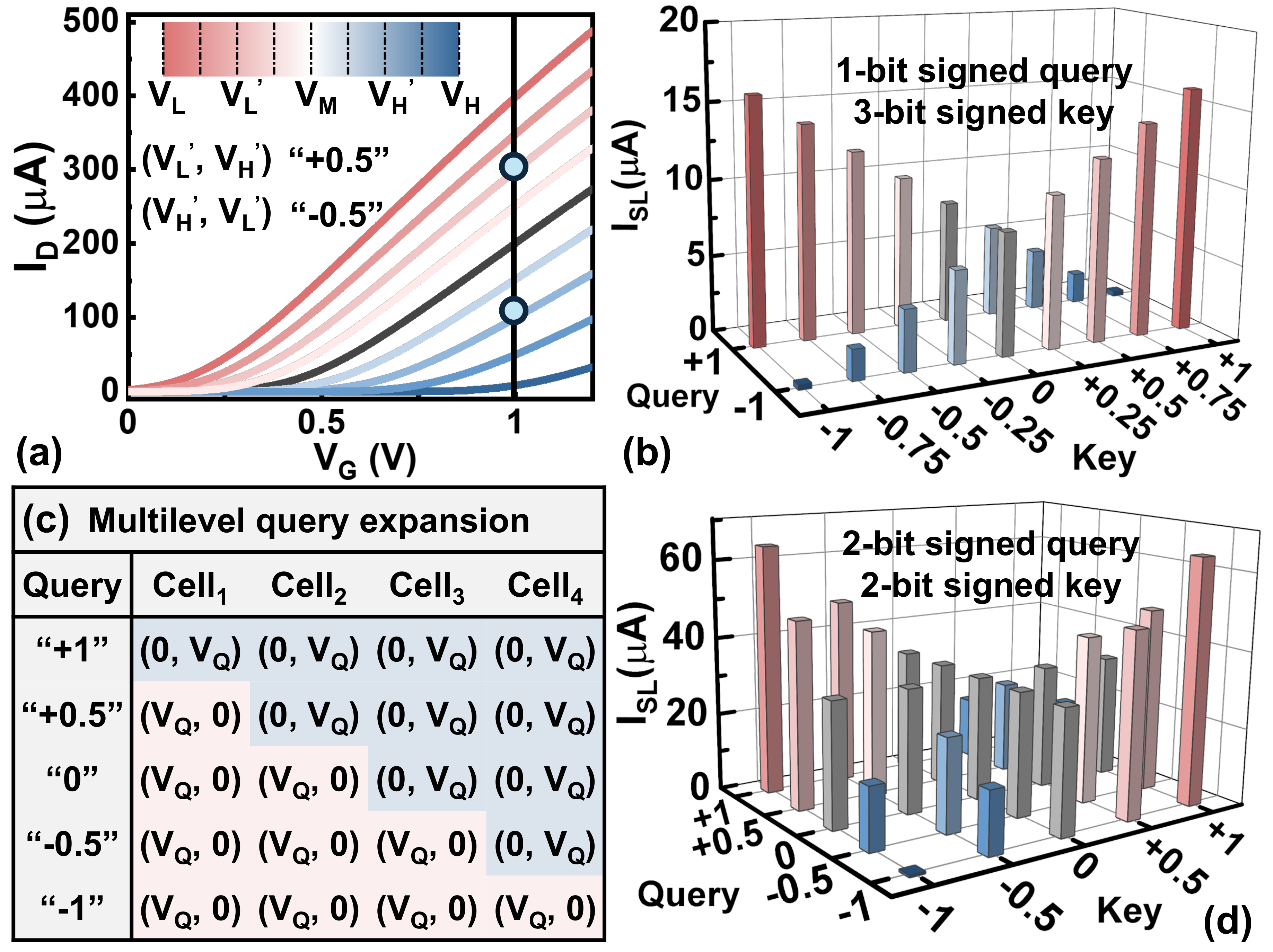}
    \vspace{-22pt}
    \caption{(a) Gradually modulated V\textsubscript{TH} of FeFET for in-place multilevel signed key expansion. (b) The computing results of local signed multiplication between 1-bit signed query and 3-bit signed key. (c) The proposed encoding method for multilevel query expansion. (d)  (b) The computing results of local signed multiplication between 2-bit signed query and 2-bit signed key.}
    \vspace{-15pt}
    \label{fig:6}
\end{figure}

\subsubsection{\textbf{FeFET-based UniCAIM cell}} 
The proposed FeFET-based UniCAIM cell is composed of two 1-transistor-1-FeFET (1T1F) units (Fig. \ref{fig:5}(a)), which can store the signed key and implement the local signed multiplication (i.e., attention score) between signed key and query. In the write stage, the signed key of "-1"/"+1’ is programmed in two FeFETs (F\textsubscript{1} and F\textsubscript{1b}) with complementary V\textsubscript{TH} states (V\textsubscript{TH1} and V\textsubscript{TH1b}) by applying relatively large V\textsubscript{P} for FE polarization switching (Fig. \ref{fig:5}(b) and (c)). Then the signed query of "-1"/"+1" is represented by complementary non-destructive read voltages at bit-lines (BL\textsubscript{1} and BL\textsubscript{1b}) with relatively small amplitudes. The signed multiplication result can be expressed through sense-line current (I\textsubscript{SL}), where the low/high I\textsubscript{SL} represents the result of "+1"/"-1", which is meticulously designed for efficient KV-cache pruning and attention computation which will be discussed in the following sections. Moreover, the Key of "0" can be represented by programming both FeFETs to the medium V\textsubscript{TH} states, which will result in medium I\textsubscript{SL} with both query inputs (Fig. \ref{fig:5}(d)).

Furthermore, by utilizing the multilevel storage capability of FeFET, the proposed FeFET-based UniCAIM cell supports  signed multi-bit storage and in-place attention computation for higher area-efficiency. With multiple V\textsubscript{P} applied to the gate of FeFET, the FE polarization can be gradually switched in succession, resulting in the continuous modulation of V\textsubscript{TH}, and the multiple V\textsubscript{TH} pairs (V\textsubscript{TH1}, V\textsubscript{TH1b}) in complementary form represent multilevel signed key states (Fig. \ref{fig:6}a). For example, the (V\textsubscript{L}', V\textsubscript{H}') and (V\textsubscript{H}', V\textsubscript{L}') can represent keys of "+0.5" and "-0.5", respectively. The signed multiplication results with 3-bit signed keys are shown in Fig. \ref{fig:6}b. Besides, the multilevel signed query expansion, including "0",  can be implemented through bitwise expansion with the proposed mapping method as shown in Fig. \ref{fig:6}(c), enabling signed multiplication with multilevel signed queries and keys (Fig. \ref{fig:6}(d)).

\begin{figure}[tb]
    \centering
    \includegraphics[width=1\linewidth]
    {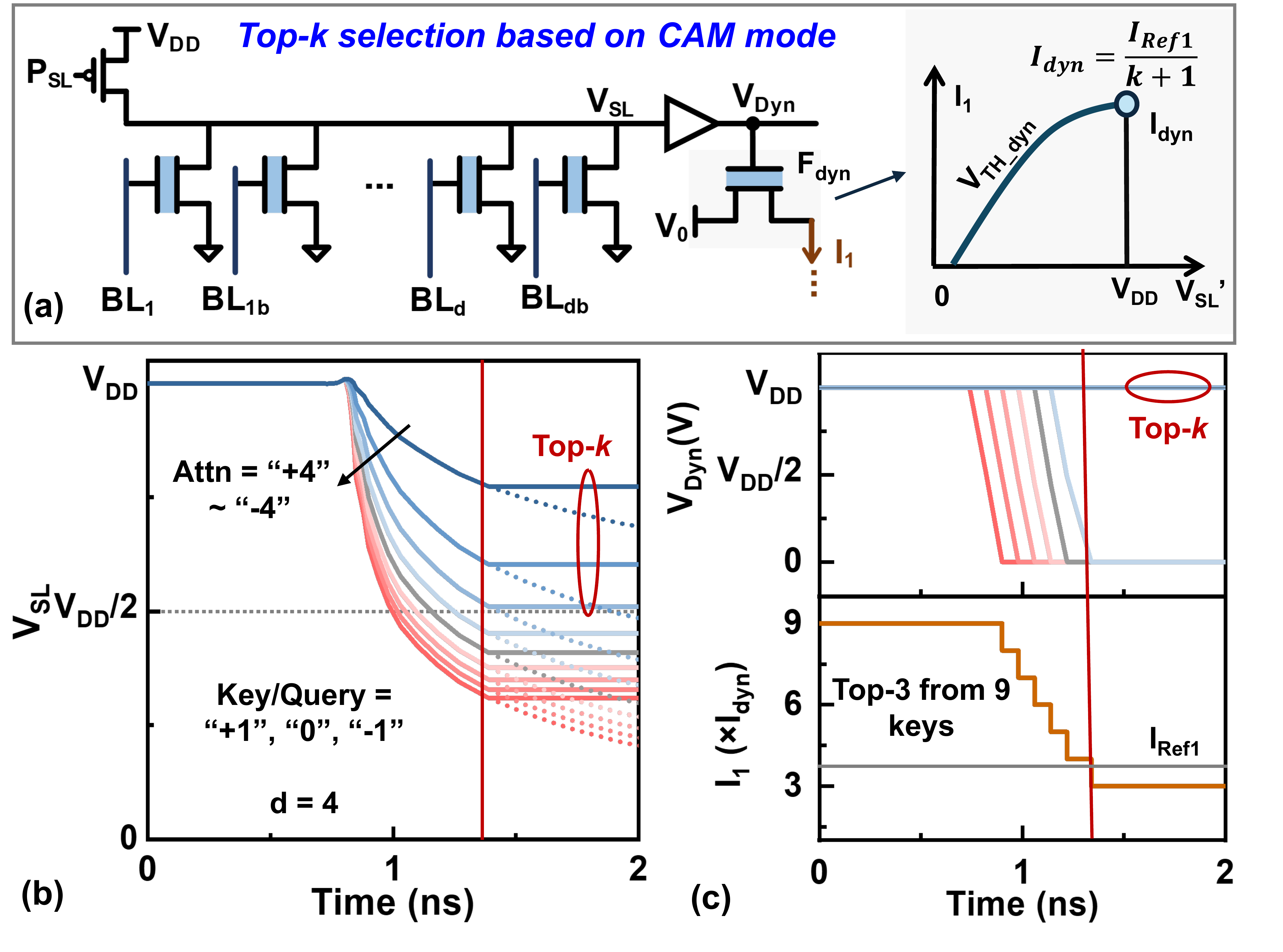}
    \vspace{-20pt}
    \caption{(a) The main circuit of CAM mode for top-\textit{k} selection. (b) The discharge speed of the sense-line (SL) is correlated with the attention scores, where the SL with higher attention discharges more slowly. (c) The sequence diagrams of top-\textit{k} selection take the example of top-3 selection from 9 keys.}
    \vspace{-15pt}
    \label{fig:7}
\end{figure}

\subsubsection{\textbf{CAM mode of UniCAIM for dynamic pruning}}
To implement the dynamic KV cache pruning, the CAM mode of UniCAIM is designed for fast and energy-efficient top-\textit{k} selection. 
As shown in Fig. \ref{fig:7}a, each row shares one precharge p-type transistor for precharging the SL and detecting circuits including one buffer and one FeFET (F\textsubscript{dyn}) with properly programmed V\textsubscript{TH}  for top-\textit{k} selection. 
In the dynamic pruning phase, all SLs are precharged to high voltage (i.e., V\textsubscript{DD}) by setting the P\textsubscript{SL} to ground, and then all BLs are prechraged to the corresponding voltages according to the proposed mapping method of the input query. 
Based on the proposed UniCAIM cell, the higher similarity between the query and the key (i.e., the higher attention score) will lead to the smaller I\textsubscript{SL}, and thus result in a slower discharge speed of the SL. For a key/query with a dimension of 4, where each state is "+1", "-1" or "0", there are 9 possible attention scores ranging from "-4" to "+4", and the corresponding SL discharge processes for different scores are shown in Fig. \ref{fig:7}b. The SLs with lower similarity will discharge to V\textsubscript{DD}/2 more quickly, resulting in V\textsubscript{Dyn} switching to ground, which turns off the corresponding F\textsubscript{dyn}. In contrast, the SLs with higher similarity maintain V\textsubscript{Dyn} at V\textsubscript{DD}, which turns on the F\textsubscript{dyn} with I\textsubscript{dyn}. By setting the I\textsubscript{Ref1} to (\textit{k}+1)I\textsubscript{dyn}, top-\textit{k} selection can be performed, as shown in the example of top-3 selection from 9 keys (Fig. \ref{fig:7}(c)). 
When \textit{k} SLs with higher similarity maintain high voltage, the accumulated I\textsubscript{1} will be less than I\textsubscript{Ref1}, causing the control signal (Ctrl\textsubscript{1}) to switch (Fig. \ref{fig:4}), which will disable the discharge process of SLs. The addresses of the top-\textit{k} selected keys are stored in the local register for exact similarity computation in subsequent steps.

The proposed FeFET-based CAM can achieve top-\textit{k} selection with $\mathcal{O}(1)$ time complexity by comparing the similarity without the need for actual computation, which improves the speed and energy-efficiency compared with conventional designs. Moreover, \textit{k} can be easily configured by programming the F\textsubscript{dyn} without the additional hardware overhead, indicating the enhanced versatility for different LLMs and contexts.

\begin{figure}[tb]
    \centering
    \includegraphics[width=1\linewidth]
    {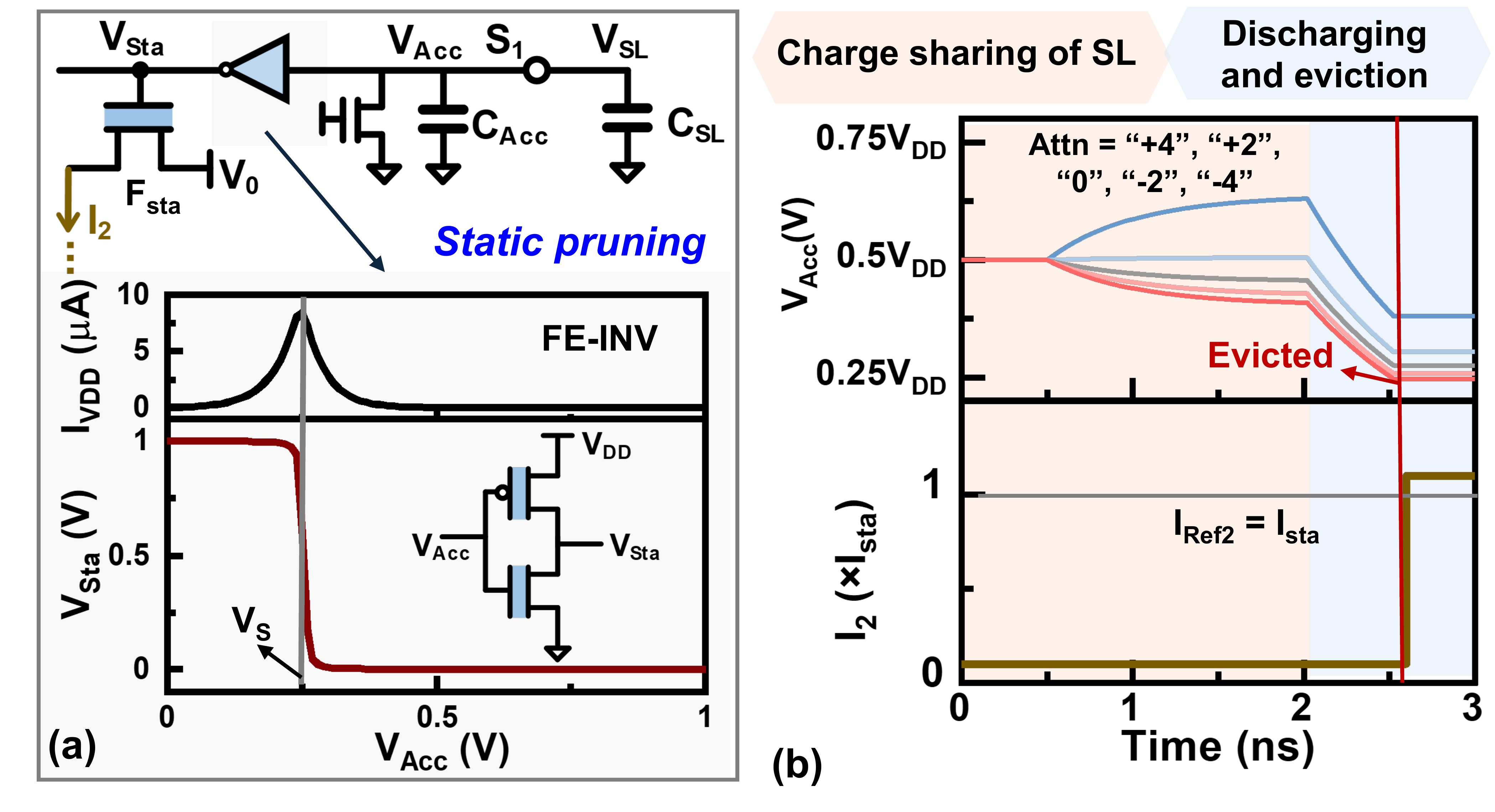}
    \vspace{-22pt}
    \caption{(a) The main circuit of charge-domain CIM mode for static pruning. (b) The process and sequence diagrams of static pruning, including charge sharing of SL for accumulating attention scores and static eviction.}
    \vspace{-13pt}
    \label{fig:8}
\end{figure}

\begin{figure}[tb]
    \centering
    \includegraphics[width=1\linewidth]
    {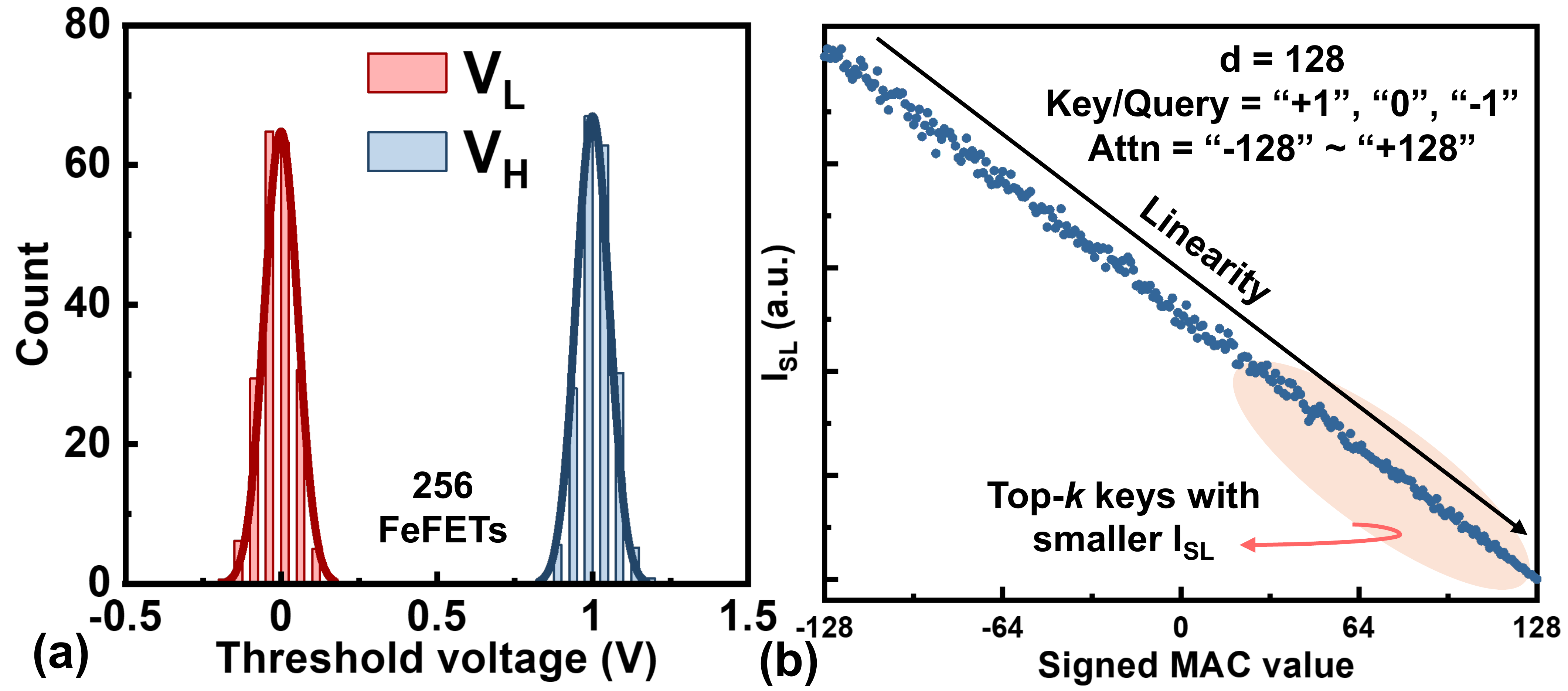}
    \vspace{-22pt}
    \caption{(a) The distribution of high/low V\textsubscript{TH} states of 128 FeFET devices. (b) The I\textsubscript{SL} is linear to the singed multiply-accumulate (MAC) value.}
    \vspace{-18pt}
    \label{fig:9}
\end{figure}

\subsubsection{\textbf{Charge-domain CIM mode of UniCAIM for static pruning}}

During the decoding stage, different tasks will generate varying lengths of tokens, which poses significant challenges to memory overhead and management. Therefore, in this work, when the number of generated tokens exceeds the reserved KV cache size (i.e., \textit{M}), the static KV cache pruning is applied. 
The charge-domain CIM is further designed for accumulating the similarity and evicting the smallest one. After implementing top-\textit{k} selection based on CAM mode, the switch S\textsubscript{1} is closed and the charge sharing occurs between SL capacitor (C\textsubscript{SL}) and accumulate capacitor (C\textsubscript{Acc}) which has accumulated the previous similarity scores (Fig. \ref{fig:8}(a)). To implement static pruning within the same operation cycle along with dynamic pruning, without the need to recharge SLs, an FeFET-based inverter (FE-INV) is designed with programmed switching voltage (V\textsubscript{S}). The SL with the smallest accumulated similarity will discharge to V\textsubscript{S} firstly through the discharge transistor, which will turn on the corresponding  F\textsubscript{sta} with I\textsubscript{sta}. Consequently, the accumulated I\textsubscript{2} is larger than I\textsubscript{Ref2} which is equal to I\textsubscript{sta}, causing the control signal (Ctrl\textsubscript{2}) to switch, which will turn off the discharge transistor (Fig. \ref{fig:8}(b)). The address of the statically selected key is stored in the local register, which will be evicted after exact similarity computation at the current generation step. Moreover, to avoid swapping memory when the stored key is evicted, the newly-added key is directly overwritten by enabling the word-line (WL) of the corresponding row and applying the appropriate V\textsubscript{P} on BLs with a single write cycle.

\subsubsection{\textbf{Current-domain CIM mode of UniCAIM for attention computation}}
To implement the exact attention computation for the top-\textit{k} selected tokens after dynamic-static KV cache pruning, the FeFET-based current-domain CIM is proposed, where the I\textsubscript{SL} is precisely quantized via the analog-to-digital converter (ADC) to calculate the accurate attention score. Fig. \ref{fig:9} shows the I\textsubscript{SL} with 128 activated FeFET-based \method~cells, which shows robust linearity with respect to the singed multiply-accumulate (MAC) value, when considering the device-to-device variation of FeFET devices with a standard deviation of 54mV \cite{cai2022energy}. Benefiting from the dynamic pruning, only the I\textsubscript{SL} of top-\textit{k} most similar keys need to be quantized by the ADCs, which can be selected through a Multiplexer (MUX). It can significantly reduce the computational overhead and the power consumption of the ADC, which is the primary source of power usage in analog CIM systems. Moreover, due to the meticulous design of FeFET-based \method~cell, the I\textsubscript{SL} is smaller when the attention score is larger, and thus the dynamically selected top-\textit{k} tokens which need to be precision computed have relatively small I\textsubscript{SL}, leading to lower energy consumption for attention computation.


\section{Experimental Results}
\label{sec:exp}

\begin{figure}[tb]
    \centering
    \includegraphics[width=1\linewidth]
    {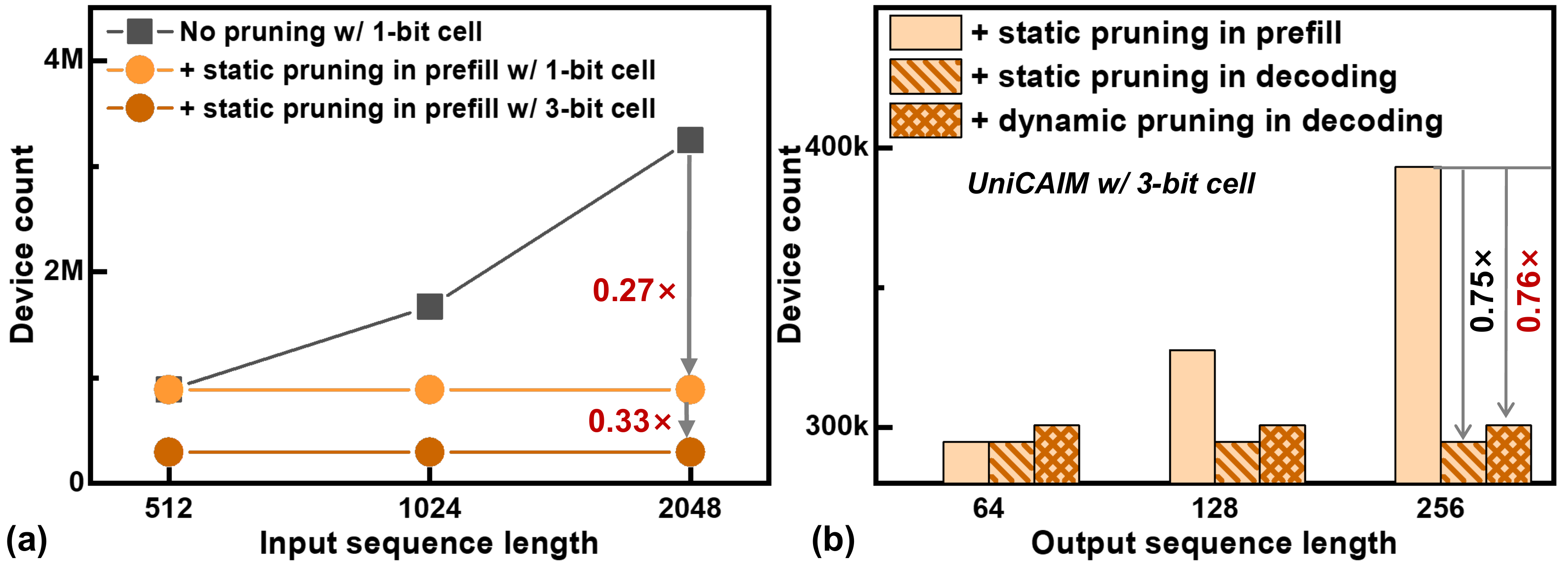}
    \vspace{-22pt}
    \caption{The required device count as (a) input sequence length and (b) output sequence length increases with different pruning conditions.}
    \vspace{-12pt}  
    \label{fig:10}
\end{figure}

\begin{figure}[tb]
    \centering
    \includegraphics[width=1\linewidth]
    {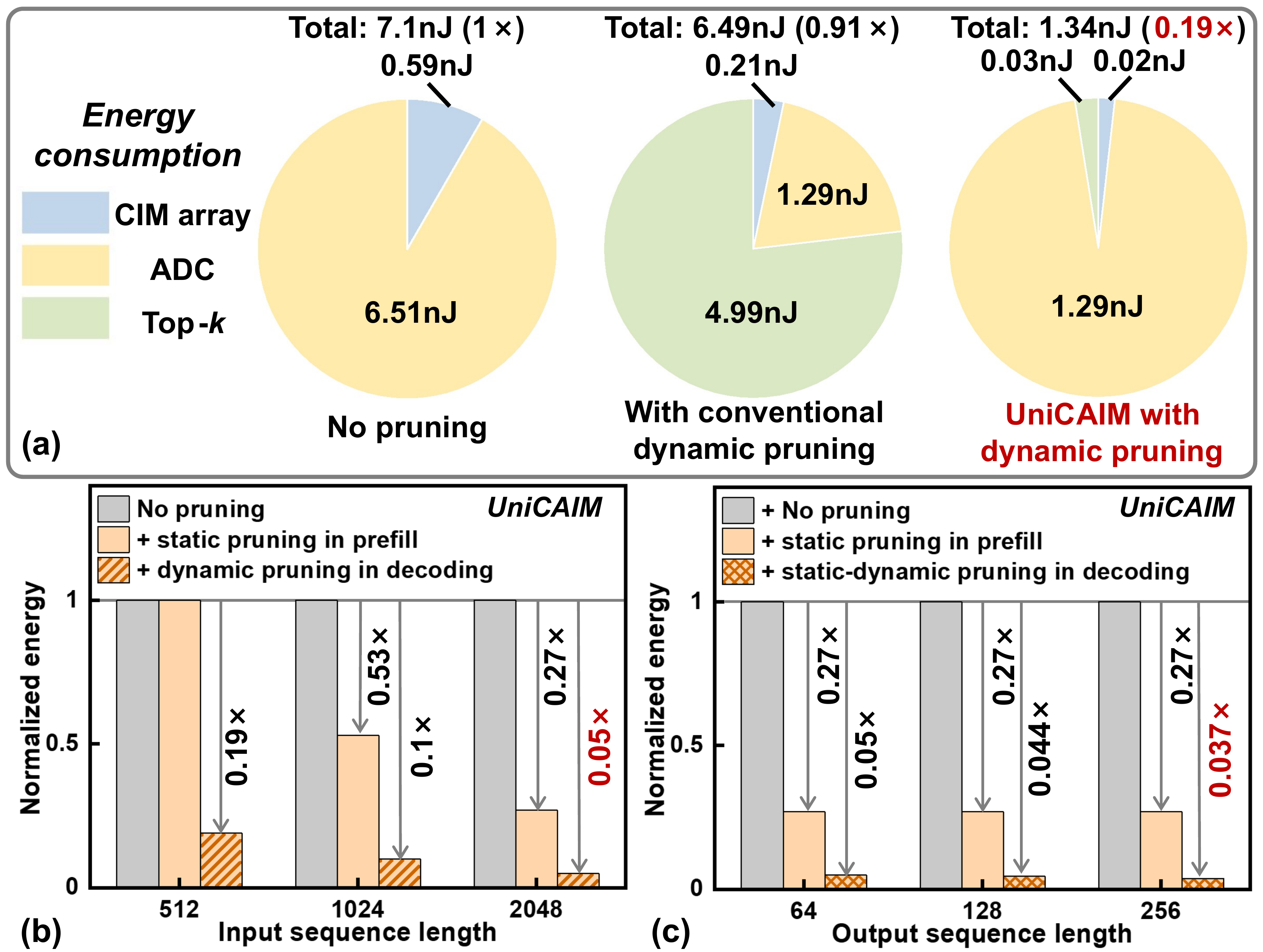}
    \vspace{-20pt}
    \caption{(a) The impact of dynamic pruning on power consumption, with a 20\% pruning ratio. The energy consumption as (b) input sequence length with an output sequence length of 64 and (c) output sequence length with an input sequence length of 2048 increases with different pruning conditions.}
    \vspace{-18pt}
    \label{fig:11}
\end{figure}

In this section, we validate and evaluate the proposed FeFET-based UniCAIM architecture for LLM acceleration. Firstly, the key performance metrics for attention including area, energy, and delay are evaluated and compared with the state-of-the-art CIM-based designs at the circuit level. Subsequently, the impact of the proposed dynamic-static KV cache pruning algorithm on the accuracy of LLM applications is evaluated.

\subsection{Circuit-level Evaluations}
\label{subsec:main_res}
The circuit-level experiments and evaluations of proposed FeFET-based \method~are carried out based on typical emerging FeFET devices with circuit simulation in HSPICE. The 45nm BSIM model \cite{zhao2006new} is used for all MOSFETs, and the Preisach ferroelectric switching model \cite{ni2018circuit} is used for FeFETs. The parasitic wire capacitance is extracted according to \cite{bhardwaj2022power}. The KV cache size contains 576 tokens, with 512 initial heavy tokens and 64 reserved tokens for decoding. Each token has a hidden dimension (i.e., \textit{d}) of 128 with a 3-bit UniCAIM cell, and the range of attention scores is “-512” to “512”, and thus a 10-bit SAR ADC is used for quantization \cite{liu201010b}.

\subsubsection{\textbf{Area}}
The static KV cache pruning during the prefill/decoding stage can reduce the KV cache size corresponding to the input/output sequence, leading to improved area efficiency, and the improvement is more significant as the input/sequence length increases due to the higher compression ratio (Fig. \ref{fig:10}). 
It indicates that static KV cache pruning is the key technique for improving area efficiency.
Moreover, benefiting from the proposed FeFET-based UniCAIM cell with \textit{in-situ} multilevel expansion capability, the area efficiency can be further improved.
Besides, the proposed CAM-based dynamic pruning circuit is highly compact, resulting in only a slight decrease in the improvement of area efficiency from 15$\times$ to 14.7$\times$ (Fig. \ref{fig:10}(b)).

\subsubsection{\textbf{Energy}}
To implement dynamic pruning, conventional designs involve approximate attention computation, followed by selecting the top-\textit{k} highest similar tokens with additional circuits, limiting the energy efficiency improvement. 
In this work, the FeFET-based CAM mode is designed for dynamic pruning, which evaluates the relative attention and selects the top-\textit{k} tokens through a single SL charge-discharge process, without the need for actually calculating attention scores and ranking them. It significantly improves energy efficiency, due to reducing the energy consumption overhead associated with ADCs, and avoiding the additional power needed for top-\textit{k} selection circuits (Fig. \ref{fig:11}(a)). 
On the other hand, the static KV cache pruning reduces the number of tokens, thereby naturally reducing energy consumption. 
Moreover, the improvement of energy-efficiency is more significant from 5.3$\times$ to 27$\times$ as the input and output sequence length increase, benefiting from the static pruning during both prefill and decoding stages (Fig. \ref{fig:11}(b) and (c)).


%

\begin{figure}[tb]
    \centering
    \includegraphics[width=1\linewidth]
    {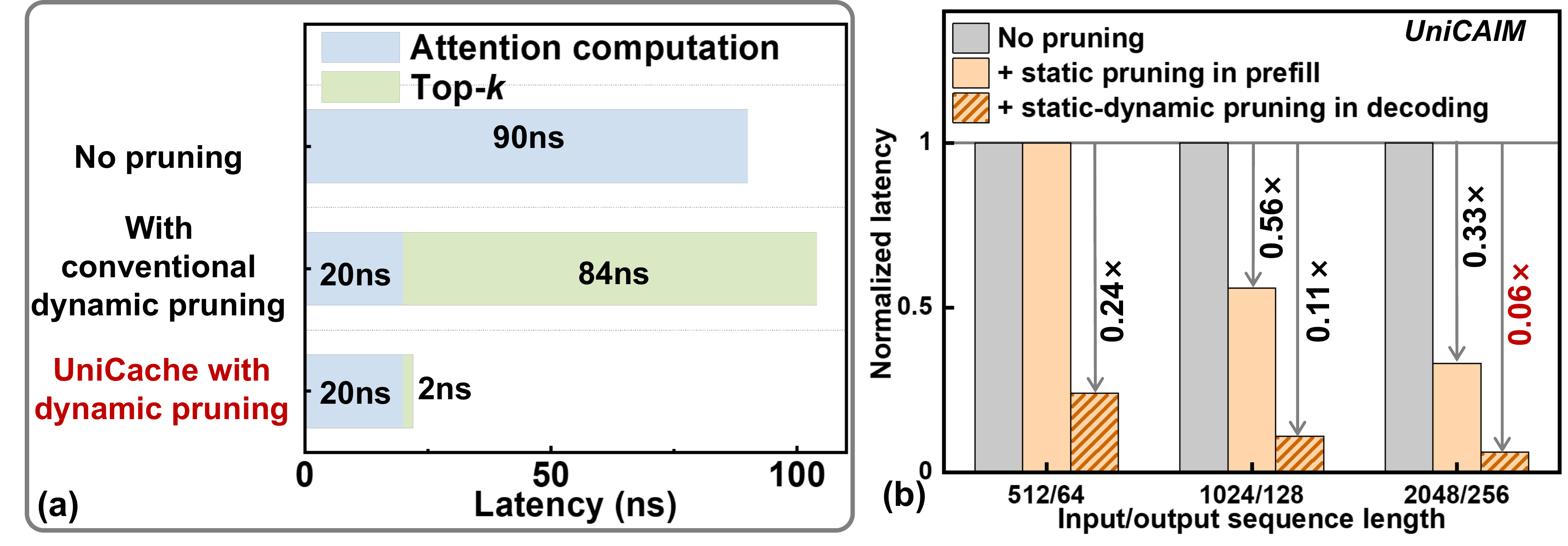}
    \vspace{-20pt}
    \caption{(a) The impact of dynamic pruning on latency, with 20\% pruning ratio, 512 input sequence length, and 64 output sequence length. (b) The comparison of latency as input and output sequence length increase with different pruning conditions.}
    \vspace{-12pt}
    \label{fig:12}
\end{figure}

{\renewcommand{\thefigure}{\Roman{figure}}
\begin{figure}[tb]
    \renewcommand{\figurename}{Table} 
    \setcounter{figure}{1} 
    \centercaption{Quantitative comparison of proposed FeFET-based UniCAIM with the state-of-the-art CIM-based LLM accelerators.}
    \vspace{3pt}
    \centering
    \includegraphics[width=1\linewidth]
    {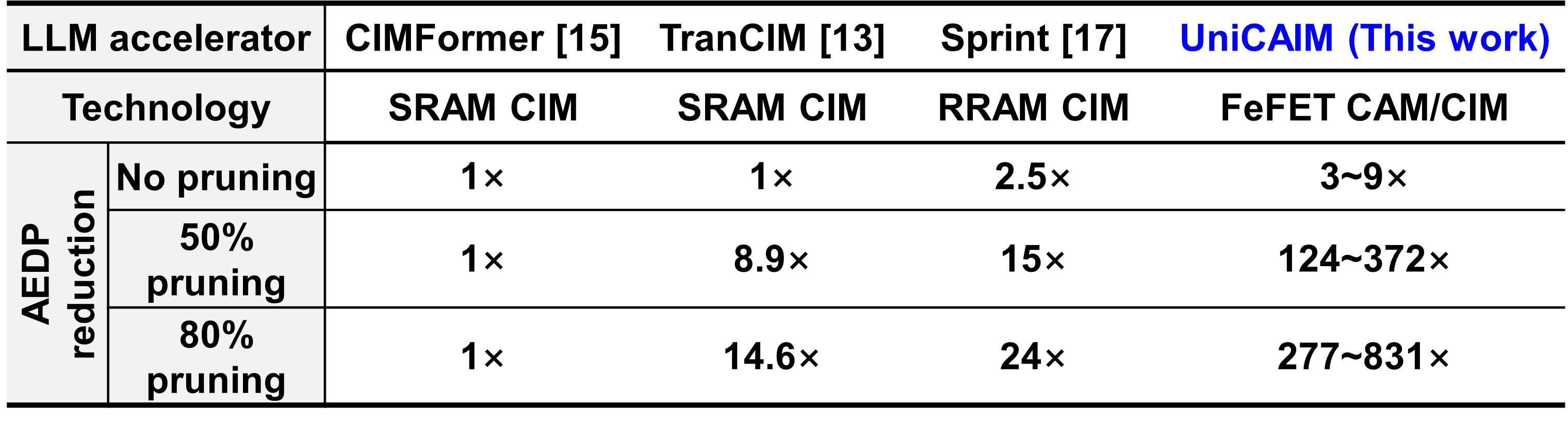}
    \vspace{-25pt}
    \label{table:2}
\end{figure}
}

\subsubsection{\textbf{Delay}}
Although CIM architecture enables parallel computation in theory, the output parallelism is constrained by the maximum number of ADCs that can be accommodated within the area and power budget of the system, which further increases the compute delay. In this work, 64 SLs are sensed in parallel with 64 ADCs.
As shown in Fig. \ref{fig:12}(a), although the conventional dynamic KV cache pruning reduces the number of attention scores that need precision quantization, the approximate attention computation is still limited by the number of ADCs. Additionally, the top-\textit{k} selection process involves complex $\mathcal{O}(n \cdot \log n)$ time complexity, which actually increases the overall computation latency. 
The proposed FeFET-based UniCAIM does not require ADC quantization during dynamic pruning and can implement the top-\textit{k} selection with $\mathcal{O}(1)$ time complexity, thereby further reducing computation latency.
Moreover, the static KV cache pruning reduces the number of attention scores that require ADC quantization, thereby reducing computation latency. 
Moreover, the speed up is more significant from 4.2$\times$ to 16.7$\times$ as the input and output sequence length increase, benefiting from the static KV cache pruning during the prefill stage and static-dynamic KV cache pruning decoding stages (Fig. \ref{fig:12}(b)).

\subsubsection{\textbf{Comparison with state-of-the-art designs}}
We use the recently reported CIM-based LLM accelerators CIMFormer \cite{cimformer}, TranCIM \cite{trancim}, and  Sprint \cite{yazdanbakhsh2022sparse} as the baselines for quantitative comparison, and we apply the same static/dynamic KV cache pruning ratio across different designs for a fair comparison. The proposed FeFET-based UniCAIM with static-dynamic KV cache pruning achieves a significant reduction in the AEDP (Table \ref{table:2}). With 1-bit FeFET-based UniCAIM cell, the AEDP is reduced by 8.2$\times$/13.9$\times$/124$\times$ and 11.5$\times$/19$\times$/277$\times$ with the pruning ratio of 50\% and 80\% compared with Sprint/TranCIM/CIMFormer, respectively. Moreover, When further utilizing a 3-bit FeFET-based UniCAIM cell, the AEDP reduction is even more pronounced, which achieves 24.8$\times$/41.7$\times$/372$\times$ and 34.6$\times$/56.9$\times$/831$\times$, indicating its great potential for long-context LLM acceleration at the edge.


\begin{figure}[tb]
    \centering
    \setcounter{figure}{12} 
    \includegraphics[width=1\linewidth]
    {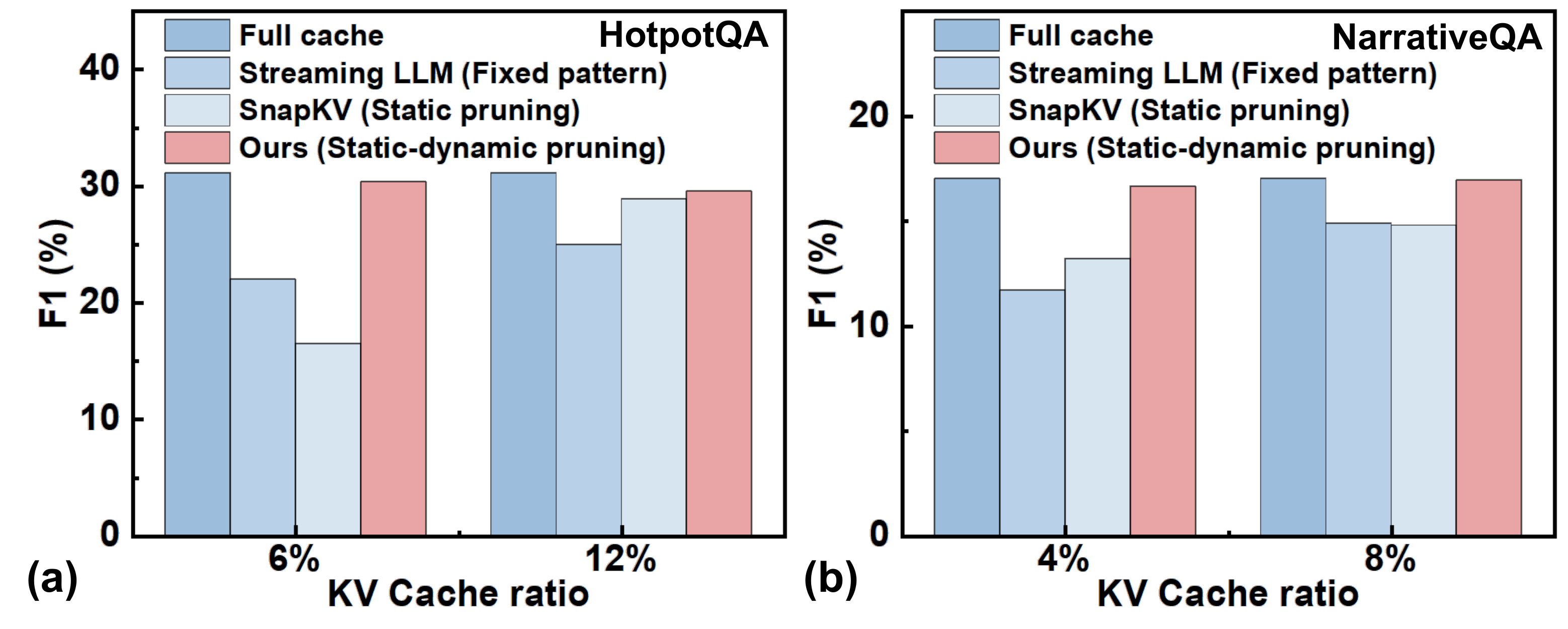}
    \vspace{-20pt}
    \caption{Accuracy evaluation of KV cache pruning on different datasets of (a) HotpotQA and (b) NarrativeQA, where F1 is a widely adopted evaluation metric that measures the similarity of the outputs to the ground-truth answers for QA and summarization tasks.}
    \vspace{-15pt}
    \label{fig:13}
\end{figure}

\subsection{Application-level Evaluation}
We evaluate our proposed static-dynamic KV cache pruning algorithm on the LongChat-v1.5-7B-32k \cite{li2023long} model on widely used long-context tasks from LongBench \cite{bai2023longbench}, including HotpotQA and NarrativeQA.
The prompt length of HotpotQA is 1.5k and the prompt length of NarrativeQA is 2.5k.
As can be observed in Fig. \ref{fig:13}, 
our proposed static-dynamic KV cache pruning algorithm does not significantly affect the accuracy of LLM  and achieves comparable accuracy with full-cache attention even under low KV cache ratios.
When compared with the recent KV cache pruning algorithms SnapKV \cite{li2024snapkv} and StreamingLLM \cite{xiao2023efficient}, our algorithm consistently achieves higher accuracy, indicating the effectiveness of our algorithm.

\section{Conclusion}
\label{sec:conclusion}
In this work, a novel FeFET-based UniCAIM architecture featuring dynamic-static KV cache pruning is proposed for sparse attention. The UniCAIM architecture can efficiently implement dynamic pruning, static pruning, and attention computation, with different CAM and CIM modes. Evaluation results at circuit and application levels show significantly reduced AEDP and high accuracy, indicating its great potential for long-context LLM acceleration at the edge.

\clearpage
\bibliographystyle{IEEEtran}
\bibliography{bib/ref}

\end{document}